# Low-phase-noise surface-acoustic-wave oscillator using an edge mode of a phononic band gap

Zichen Xi[1,2], Joseph G. Thomas[1,3], Jun Ji[1,2], Dongyao Wang[4,5], Zengyu Cen[4,5], Ivan I. Kravchenko[6], Bernadeta R. Srijanto[6], Yu Yao[4,5], Yizheng Zhu[1,3], and Linbo Shao[1,2,7,*]

[1]Bradley Department of Electrical and Computer Engineering, Virginia Tech, Blacksburg, VA 24061, USA
[2]Center for Quantum Information Science and Engineering (VTQ), Virginia Tech, Blacksburg, VA 24061, USA
[3]Center for Photonics Technology, Virginia Tech, Blacksburg, VA 24061, USA
[4]School of Electrical, Computer and Energy Engineering, Arizona State University, Tempe, AZ, 85281 USA
[5]Center for Photonic Innovation, Arizona State University, Tempe, AZ, 85281 USA
[6]Center for Nanophase Materials Sciences, Oak Ridge National Laboratory, Oak Ridge, TN 37830 USA
[7]Department of Physics, Virginia Tech, Blacksburg, VA 24061, USA
*Correspondence to: shaolb@vt.edu

**Abstract**
Low-phase-noise microwave-frequency integrated oscillators provide compact solutions for various applications in signal processing, communications, and sensing. Surface acoustic waves (SAW), featuring orders-of-magnitude shorter wavelength than electromagnetic waves at the same frequency, enable integrated microwave-frequency systems with much smaller footprint on chip. SAW devices also allow higher quality ($Q$) factors than electronic components at room temperature. Here, we demonstrate a low-phase-noise gigahertz-frequency SAW oscillator on 128°Y-cut lithium niobate, where the SAW resonator occupies a footprint of 0.05 mm². Leveraging phononic crystal bandgap-edge modes to balance between $Q$ factors and insertion losses, our 1-GHz SAW oscillator features a low phase noise of -132.5 dBc/Hz at a 10 kHz offset frequency and an overlapping Hadamard deviation of $6.5 \times 10^{-10}$ at an analysis time of 64 ms. The SAW resonator-based oscillator holds high potential in developing low-noise sensors and acousto-optic integrated circuits.

## I. Introduction

Oscillators, which generate periodically alternating signals with a stable frequency and phase, play crucial roles in modern telecommunication, metrology, and sensing systems. Complying constraints in size, weight, and power (SWaP), miniaturized microwave-frequency oscillations are developed using integrated electronic circuits [1], microwave photonics [2-8], optomechanical devices [9-11], chip based atomic clocks [12,13] and acoustic-wave devices [14-22]. Recently, microwave integrated photonics demonstrates chip-based oscillators with ultralow phase noise [2-4], despite integration of all needed components, including laser [23,24], photodiode [25], optical resonators, optical amplifier, and feedback control circuits, on a single chiplet remains challenging.

Meanwhile, integrated microwave acoustics [26] with electronic circuitry enables simpler oscillator architectures with competitive performance metrics. Compared to electronic counterparts, acoustic-wave resonators [27-33] provide higher quality ($Q$) factors in smaller footprints, leveraging the orders-of-magnitude smaller velocity of acoustic waves in solids than electromagnetic waves. Phase velocities of acoustic waves in solids are typically within a range of a few kilometers per second resulting in hundreds of nanometers to micron scale feature sizes for GHz-frequency acoustic-wave devices, which are nanofabrication friendly with research facilities and industrial foundries. The mechanical nature of acoustic waves results in immunity to electromagnetic noise and crosstalk in a compact package. In addition, acoustic waves can be efficiently bidirectionally transduced to electrical signals via the piezoelectric effect of materials, such as, lithium niobate (LN) [18,29,32], quartz [15,31,33], gallium nitride [14,19] and aluminum nitride [16,17,20,21].

Acoustic-wave oscillators employ either acoustic-wave delay lines or resonators to achieve low phase noise. Generally, a longer phase delay and a lower insertion loss of the acoustic-wave device will lead to a lower phase noise [34]. High-$Q$ acoustic-wave resonators can provide equivalent delay times with larger frequency spacing (free spectral range) between resonant modes and smaller footprint than acoustic-wave delay lines. For example, the delay time of an acoustic-wave resonator with a $Q$ factor of 2,000 at 1 GHz is equivalent to a 2-mm-long acoustic delay line on lithium niobate. Integrated acoustic-wave resonators in suspended [28,30,32], thin-film-on-bulk [29], and surface [27,31] acoustic-wave device architectures have been developed with frequencies from MHz to sub-THz [32]. Their $Q$ factors range from several hundreds to tens of thousands with frequency-$Q$ ($fQ$) product reaching $10^{13}$ at room temperature. At cryogenic temperature, $Q$ over ten billion ($10^{10}$) and $fQ$ over $10^{20}$ have been achieved in a nano acoustic resonator [35].

Here, we demonstrate a microwave-frequency low-phase-noise surface acoustic wave (SAW) oscillator using a phononic crystal (PnC) resonator on LN platform. We leverage a bandgap-edge mode of the PnC resonator to trade off $Q$ factor against insertion loss to achieve the low phase noise. Our 1-GHz oscillator features an output power of 2.71 mW (4.33 dBm), phase noise of -132.5 dBc/Hz at a 10-kHz offset frequency, and a minimum Hadamard deviation (long-term stability) of $6.5 \times 10^{-10}$ at analysis time τ ~ 64 ms. The SAW oscillators with small mode volumes are promising candidates in developing oscillator-based sensors.

## II. Device design principle and fabrication

The phase noise, $PN$, of a resonator-based oscillator can be estimated by the Leeson's formula [34],

$$PN \approx -174 - P_c + NF_{LNA} + IL + 20\, log_{10}\left(\frac{f_0}{2f_m Q}\right) \qquad (1)$$

where $PN$ is the phase noise in dBc/Hz, $P_c$ is the carrier power at the amplifier output in dBm, $IL$ is the insertion loss of the loop in dB, $NF_{LNA}$ is the noise figure of the low-noise amplifier (LNA) in dB, $f_0$ is the oscillation frequency, $f_m$ is the offset frequency, and $Q$ is the quality factor of the resonator. The $P_c$ and $NF_{LNA}$ is determined by the performance characteristics of the LNA. This work focuses on the acoustic resonator, which mainly determines the insertion loss $IL$, resonant frequency $f_0$, and $Q$ in the Leeson's formula.

Towards a low-phase-noise oscillation, the Leeson's formula suggests that the product of transmission $T$ and $Q$ as the figure of merit for the resonator. This motivates our usage of the bandgap-edge mode of a PnC resonator. Compared to the previous PnC resonant modes [27] locating at frequencies deep in the bandgap, the bandgap-edge mode locates near the bandgap-edge frequencies. While the bandgap-edge mode is still a confined mode maintaining a high $Q$ factor, the acoustic waves at bandgap-edge frequencies can propagate deeper into the PnC mirror and result in a much higher external coupling efficiency, i.e., transmission $T$.

Our PnC resonator (Fig. 1) is fabricated on a 128°Y-cut LN substrate with SAW propagating along crystal X axis. Lithium niobate features large piezoelectricity, low acoustic-wave propagation loss at GHz frequencies, and mature nanofabrication processes [36]. The 128°Y-cut X-propagating configuration shows high electromechanical coupling efficiency ($k^2$) and a low acoustic-wave diffraction loss. We use the black LN to mitigate the pyroelectric issue during fabrication processes. We define the PnC resonator by a series of etched grooves (Fig. 1(a)), which are patterned by electron beam lithography (EBL) and etched by reactive ion etching (RIE) using argon gas. The target depth of the grooves is 100 nm. The metal layer of interdigital transducers (IDTs) is patterned by EBL aligned to etched markers, which centers the electrodes between two etched grooves (Figs. 1(b) and 1(c)). We deposit aluminum of 100 nm using an electron beam evaporator followed by a lift-off process. To minimize the perturbation to the resonant mode by the electrodes, we choose aluminum over gold for its higher conductivity over density ratio.

We design the resonator by varying the period and width of etched grooves in different regions (Fig. 2(a)). The center segment (labeled Segment A in Fig. 2(a)) is with grooves of period 2 μm and gradually transits to segments (labeled Segment B, C in Fig. 2(a)) of period 1.96 μm. Segments B and C further transit to unetched free surface by tapering the width of etched grooves. These gradual transitions reduce the scattering loss of acoustic waves into the substrate and improve $Q$ factor of acoustic resonant modes. We make the number of grooves in Segment B less than that in Segment C to optimize the external coupling efficiency from the IDT to the resonator modes. The optimized numbers of periods in PnC segments and tapers are 10 periods in Taper A, 60 periods in Segment B, 30 periods in Taper B, 10 periods in Segment A, and 100 periods in Segment C.

A pair of IDTs are used to excite and detect acoustic waves. The side IDT (labeled IDT 1 in Fig. 2(a)) is positioned outside the groove region and designed to closely match the external 50 Ω impedance for efficient conversion between electrical and acoustic-wave domain. The central IDT (labeled IDT 2 in Fig. 2(a)) is positioned inside the groove region for effectively coupling to the confined acoustic-wave mode. We place electrodes of IDT 2 at center of unetched part between etched grooves to optimally overlap them with our interested acoustic mode profiles. Due to the resonant enhancement, less electrodes are needed inside the Segment A region.

The simulated band structure diagram (Fig. 2(b)) shows a bandgap from 963 to 1,002 MHz (982 to 1,024 MHz) in Segment A (Segments B and C) structures. Upper modes of Segment A are within the bandgap of Segments B and C. Due to the short length of Segment A, only a few resonant modes can be formed on the upper band of Segment A.

### III. Characterization of surface acoustic wave resonator

We characterize our SAW resonator by measuring $S$ parameter spectra (Figs. 2(c) and 2(d)). The $S$ parameter measurement is calibrated to the IDT electrode contact pads. We perform numerical simulations of eigenmodes (Figs. 2(e)-(g)) using COMSOL Multiphysics and experimentally measure mode profiles (Figs. 2(h)-(j)) using our in-house optical vibrometer [37], which features a detectable frequency up to 20 GHz and a displacement sensitivity of 0.1 pm.

Our SAW resonator supports three resonant modes. Modes 1 and 2 (marked by the two red dots in Fig. 2(b)) are well confined modes within the bandgap of Segments B and C (Figs. 2(c), 2(e) and 2(f)). The interested bandgap-edge mode, labeled Mode 3 (marked by the dark blue dot in Fig. 2(b)), is at the frequency of upper bandgap edge of Segments B/C (Figs. 2(c), 2(d) and 2(g)). In the $S$ parameter spectra measurements, we connect Ports 1 and 2 of a vector network analyzer (Keysight P5004A) to the IDTs 1 and 2, respectively. We observe that within the bandgap frequency ranging from 984 MHz to 1,026 MHz (highlighted by the light blue in Fig. 2(c)), transmission $S_{21}$ is suppressed to the level of -50 dB. This bandgap frequency range is in good agreement with the simulated results (Fig. 2(b)).

The measured $S$ parameter spectra (Fig. 2(c)) clearly show a transmission $S_{21}$ peak and a reflection $S_{22}$ dip at 1,005.59 MHz, corresponding to Mode 1 (the fundamental mode). Mode 1 has a measured loaded $Q$ of ~500 with a transmission $S_{21}$ of -39 dB (0.012%). The simulated eigenmode profile of Mode 1 (Fig. 2(e)) agrees with the measured displacement profile (Fig. 2(h)). We note that the optical vibrometer does not measure the eigenmode profiles, but displacement profiles excited by the side IDT 1 using a continuous microwave source at the corresponding resonant frequencies. Thus, large displacements near the side IDT 1 are observed.

The simulated eigenmode profile of Mode 2 (Fig. 2(f)) indicates that it is a second-order mode within the bandgap frequencies. We note that Mode 2 has little overlap with central IDT 2, it is not clearly observed in the $S$ parameter spectra (Fig. 2(c)). On the other hand, when excited by the side IDT 1, Mode 2 is observed by our optical vibrometry (Fig. 2(i)).

The bandgap-edge mode (Mode 3) exhibits a transmission $S_{21}$ peak and a reflection $S_{22}$ dip at 1,025.98 MHz, which is near the upper bandgap-edge frequency of Segments B/C (Fig. 2(c)). The bandgap-edge mode has a measured higher loaded $Q$ of 2,800 with a significantly higher transmission of -20 dB (1.0%) [Fig. 2(d)] than Mode 1. The bandgap-edge mode is thus preferred to build a lower-phase-noise oscillator. While Modes 1 and 2 are confined in Segment A and its nearby taper regions, the measured profile of this bandgap-edge mode extends to Segment B and C (Figs. 2(g) and 2(j)). The profile of the bandgap-edge mode has two nodes near both boundaries of Segment A and shows a large field at central IDT 2 region which allows an efficient coupling.

We noted that loaded $Q$ of the bandgap-edge mode (Mode 3) is higher than that of the fundamental mode (Mode 1). This observation can be explained by two reasons. (1) The better confinement of the fundamental mode within the bandgap leads to a higher $Q/V$, where $V$ is the mode volume, not necessarily a higher $Q$. As shown in Figs. 2(e) and 2(g), the bandgap-edge mode has a much larger $V$ than that of the fundamental mode. (2) The IDT 2 provides a stronger coupling to the fundamental mode than that to the bandgap-edge mode, as indicated by a larger dip in the reflection $S_{22}$ spectrum (Fig. 2(c)). The measured loaded $Q$ is contributed by both intrinsic $Q$ and mode coupling rate.

### IV. Stable acoustic-wave oscillation and characterization

We achieve SAW oscillation using a positive feedback loop (Fig. 3(a)), which consists of our acoustic resonator, a LNA (Mini-circuits, ZKL-33ULN-S+), a microwave attenuator, a phase shifter (RF-LAMBDA RFPSHT0002W1), and a coupler (Mini-circuits, ZFDC-10-5-S+). The self-oscillation occurs at a frequency, where the loop phase delay is an integer number of $2\pi$, and the losses are fully compensated by the gain provided by LNA.

We characterize our oscillator by measuring output signals of the oscillator coupled out from the coupler. Four equipment configurations are employed (Insets (1)-(4) in Fig. 3(a)). A sinusoidal waveform (Fig. 3(b)) with an amplitude of ~0.5 V and a period of ~1 ns is captured by the oscilloscope (Rohde & Schwarz RTO6) (Inset (1) in Fig. 3(a)). The amplitude of oscillator output signal is determined by the saturation power of the LNA and coupling ratio of the coupler. Captured by a spectrum analyzer (Keysight, P5004A, Spectrum analyzer mode), the frequency spectrum of the oscillator output (Fig. 3(c)) shows a maximum power of 4.33 dBm at 1,025.88 MHz, which matches the resonant frequency of the bandgap-edge mode. Higher order harmonics are observed with power of -16.9 dBc; these higher order harmonics are due to the nonlinear saturation of LNA.

We employ an in-phase/quadrature (I/Q) demodulator to characterize the phase noise of our oscillator. An ultralow-phase-noise microwave generator (Keysight N5183B with low phase noise option) with a spec (typical) phase noise of -139 (-146) dBc/Hz (at 1 GHz carrier frequency) at 10 kHz offset is used as a reference local oscillator (Inset (3) in Fig. 3(a)). Experimentally, we tune the phase shifter to maximize the output power, and this manipulation also minimizes the phase noise, experimentally. We observed that the phase noise was minimized after we introduced a 3 dB attenuator into the loop. We suspect that the noise performance degradation of the LNA at deep gain saturation region results in the phase noise improvement by the additional attenuation. Our oscillator reaches a phase noise of -132.5 dBc/Hz (at ~1,026 MHz carrier frequency) at a 10 kHz offset (Fig. 3(d)), which is comparable to commercial electronic oscillators but 20 to 30 dB worse (with oscillation frequency normalized) than the state-of-the-art integrated microwave photonic oscillators [2,3]. We note that the estimation by Leeson's formula: taking $P_c = 19\ dBm$, $IL = 26.2\ dB$ (SAW resonator: 20 dB, cables in the loop: 2 dB, attenuator: 3 dB, coupler: 1.2 dB) and $NF_{LNA} = 0.5$ dB into Eq (1), it suggests a phase noise of -141 dBc/Hz at 10 kHz offset, which is lower than our measured results. We suspect that the difference between measured and estimated phase noise is caused by the underestimation of the noise figure of LNA in the gain saturation region, where its noise figure could be significantly higher than the specification value for small input signals.

We further characterize the frequency stability of our oscillator over a time scale from 10 μs to 10,000 s (Fig. 4). The long-time measurement up to 60,000 seconds (17 hours) uses a frequency counter (Keysight 53230A, 200 M points memory) (Inset (4) in Fig. 3(a)). We note that our results in Fig. 4 are in an open lab environment without any temperature feedback. The measured frequency shifts (Fig. 4(e)) match our lab temperature – the building air conditioning is turned off at night and back on in morning. Due to the resolution limit of the frequency counter, the measurements from 10 μs to 100 s use the I/Q demodulation (Inset (3) in Fig. 3(a))). Constrained by the storage length (200 M points in total) of the oscilloscope, the sampling rates are adjusted accordingly in measurements with different time lengths. The measurements with length of 0.1, 2, 100 seconds, I/Q sampling rates are set to 500, 25, 0.5 MSa/s. The minimum overlapping Hadamard deviation of our oscillator is $6.5 \times 10^{-10}$ at analysis time $\tau \sim 64$ ms, demonstrating an outstanding frequency stability. There is a 150 Hz frequency modulation noise with peak-to-peak amplitude ~ 40 Hz (Fig. 4(b)). This frequency modulation noise also causes the fluctuations in the range of 1 ~ 20 ms in the overlapping Hadamard deviation (Fig. 4(a)). We suspect that these noises are introduced by the DC power supply, as we observed a voltage ripple peak at the same 150 Hz (See details in Appendix A).

In addition, we characterize the temperature coefficient of frequency (TCF) of our device using a Peltier (thermoelectric) module placed under the chip with the temperature ranging from 15 to 42 °C (Fig. 5). We compare the TCF of the oscillator and the passive SAW resonator. The frequency counter is used to measure the oscillating frequency, and the vector network analyzer is used to extract the mode resonant frequency by fitting the transmission peaks. We measured a TCF of -70 ppm/°C for both passive resonator and active oscillator cases. This TCF is at a similar level compared to other surface or thin-film acoustic wave devices on LN platform [38-41]. We note that LN has anisotropic temperature coefficients and engineering of acoustic-wave modes on different LN crystal orientations could be performed to either reduce the TCF for stable oscillation or enhance the TCF for sensor development.

### V.     Conclusion and outlook

In conclusion, we demonstrate a low-phase-noise oscillator at 1 GHz based on a SAW resonator. Compared to previous SAW PnC resonators, our resonator features higher $TQ$ product using the bandgap-edge mode, resulting in a low phase noise of our oscillator. By integrating an on-chip thermal-acoustic phase modulator [42] and LNA dies, all components of the SAW oscillators could be compactly packaged together for SWaP-constrained applications.

The oscillating frequency of our SAW oscillator is scalable by geometrically scaling the PnC resonator design. Geometrical scaling of PnC resonators have been demonstrated from 0.5 to 5 GHz [27]. Scaling towards lower frequencies will be limited by the possible groove depth due to the poor selectivity in LN etching using polymer resist. Scaling towards higher frequencies could be limited by the feature size of the grooves. Without changing the design, a 10-GHz version of our PnC resonator will lead to a feature size of 100 nm, which is one quarter of the acoustic wavelength (0.4 μm) at 10 GHz. The current electron beam lithography is challenging to pattern better than 100 nm resolution on bulk LN substrate due to the severe charging and proximity effects. Meanwhile, thin-film LN could mitigate these issues by using silicon substrate, which is slightly doped and weakly conductive, and PnC resonators beyond 10 GHz is possible. In addition, materials with faster acoustic-wave speeds, such as aluminum nitride on diamond [43,44], could reach higher frequencies feature sizes within nanofabrication capabilities.

Beyond acoustic-wave devices, our SAW oscillators could be integrated with other optical, electro-optic, acousto-optic components to form a large-scale multi-physics integrated circuits for applications in microwave signal processing, sensing, and THz technologies.

**Appendix A: Waveform and spectra of DC power supply**

The ripples of a DC power supply could be coupled into oscillation and induce frequency noises through the LNA. The waveform (Fig. 6(a)) and frequency spectrum (Fig. 6(b)) of the DC power supply (Rigol 832A, 2.8 V input voltage with 0.5 A current limit) are captured and calculated by an oscilloscope (Rohde & Schwarz RTO6). We observe several peaks in the spectrum: most are the AC frequency (60 Hz) and its higher-order harmonics. The notable peak at 153 Hz, which is likely induced by the internal circuits of the DC power supply, matches the frequency modulations shown in Fig. 4(b). We note that this DC power supply overall has the lowest ripple voltage among the ones available in our lab.

**Acknowledgement**

We thank Prof. J. Walling for microwave instrumentation, Dr. S. Ghosh and Dr. M. Benoit for probe station in the cleanroom for quick tests. Device fabrication was conducted as part of a user project (CNMS2022-B-01473, CNMS2024-B-02643) at the Center for Nanophase Materials Sciences (CNMS), which is a DOE Office of Science User Facility. This work is supported by 4-VA Pre-Tenure Faculty Research Award, Virginia Tech FY23 ICTAS EFO Opportunity Seed Investment Grant, 2023 Ralph E. Powe Junior Faculty Enhancement Awards by Oak Ridge Associated Universities (ORAU), and the Defense Advanced Research Projects Agency (DARPA) OPTIM program under contract HR00112320031. The views and conclusions contained in this document are those of the authors and do not necessarily reflect the position or the policy of the Government. No official endorsement should be inferred. Approved for public release; distribution is unlimited.

**Conflict of Interest**

The authors have no conflicts to disclose.

**Author contributions**

Z.X. and L.S. designed, fabricated, and characterized the SAW oscillator with contributions from all other authors. J.G.T. and Y.Z. designed and built the in-house optical vibrometer, and measured mode displacement profiles. Y.Y., Z.C., and D.W. characterized the oscillator performance at ASU. I.I.K. and B.R.S. and J.J. contribute to nanofabrication and process optimization. L.S. performed the SEM imaging. Z.X. drafted the manuscript with revisions from all other authors. L.S. supervised the project.

**Data Availability**

The supporting data for this article are openly available from Figshare:
https://doi.org/10.6084/m9.figshare.28235840

## Figures

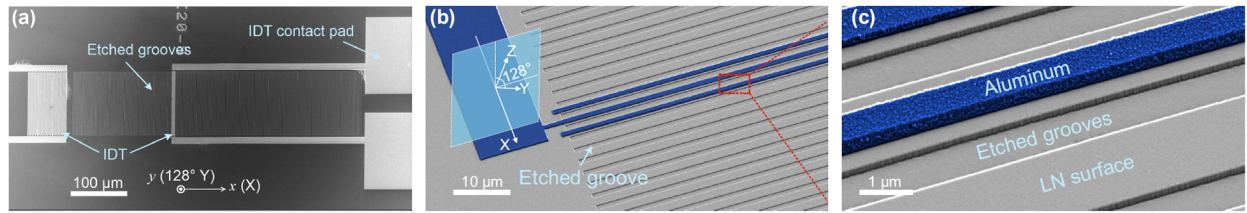

**FIG. 1.** (a) Scanning electron microscopy (SEM) images of the surface acoustic wave (SAW) resonator. The SAW resonator consists of a series of etched grooves and a pair of interdigital transducers (IDTs). One IDT is placed outside the etched groove region, which is designed to closely match the 50 Ω for efficient SAW excitation. The other IDT is placed inside the etched groove region for effectively coupling to the SAW modes. (b), (c) Zoomed in view of central region of the SAW resonator. Electrodes are placed on unetched surface between two grooves. *x, y, z* indicates device spatial coordinates; X, Y, Z indicates crystalline axis of lithium niobate. The metal in (b)(c) is false colored for better illustration.

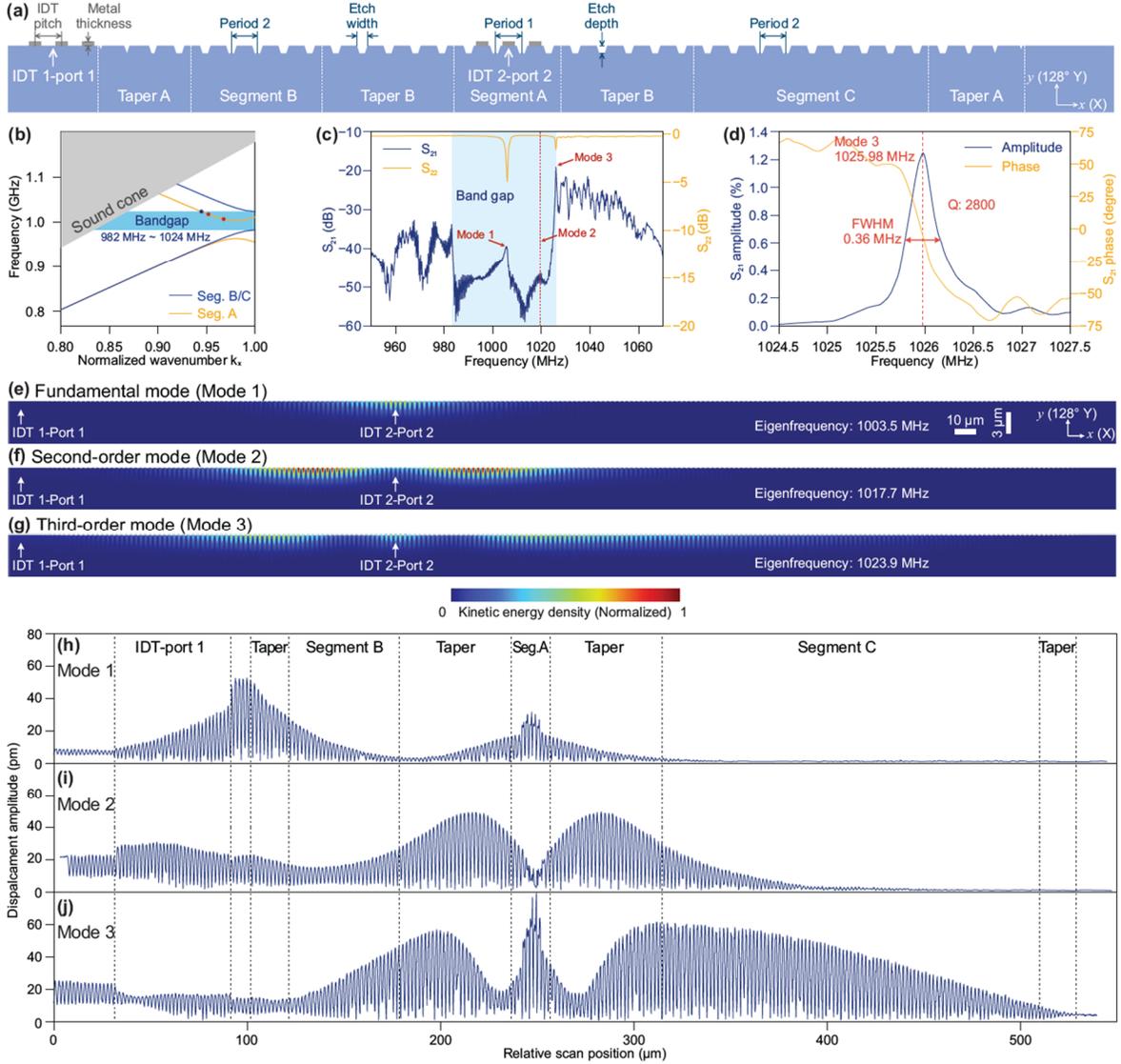

**FIG. 2.** Design and characterization of the surface acoustic wave (SAW) resonator. (a) Cross-section schematic of the SAW resonator. The SAW resonator is defined by a series of grooves with different periods in different segments. Device parameters are Period 1 is 2.00 μm, Period 2: 1.96 μm, etch width: 0.65 μm, etch depth: 100 nm, interdigital transducer (IDT) pitch: 1.91 μm, and metal thickness: 100 nm. (b) Simulated phononic band (PnC) structure of Segments A, B and C. Seg. A form a bandgap ranging from 963 to 1002 MHz; Segs. B/C form a bandgap ranging from 982 to 1024 MHz. Two SAW modes are inside PnC bandgap of Segs. B/C (marked by two red points). One SAW mode is at upper bound of bandgap of Segs. B/C (marked by dark blue point). (c) Measured transmission $S_{21}$ and reflection $S_{22}$ spectra of the SAW resonator. Two modes are observed, labeled as Mode 1 and 3. Mode 1 is inside the PnC bandgap of Segs. B/C. Mode 3 is at the upper bound of the PnC bandgap of Segs. B/C. Measured bandgap ranges from 984 to 1026 MHz (highlighted by light blue). (d) Measured amplitude and phase of transmission $S_{21}$ spectra near frequency of Mode 3. Mode 3 is centered at 1025.98 MHz with full width half maximum (FWHM) of 0.36 MHz, resulting in a quality ($Q$) factor of 2,800. (e), (f), (g) Simulated SAW eigenmode profiles supported by the SAW resonator. Mode 1 and 2 are spatially confined within Seg. A and its nearby taper regions. Mode 3 with a larger mode profile extends to Segs. B and C. Color bar indicates normalized kinetic energy density. (h), (i), (j) Measured displacement profiles of SAW modes using our in-house optical vibrometer. SAW modes are excited by applying continuous wave to IDT 1.

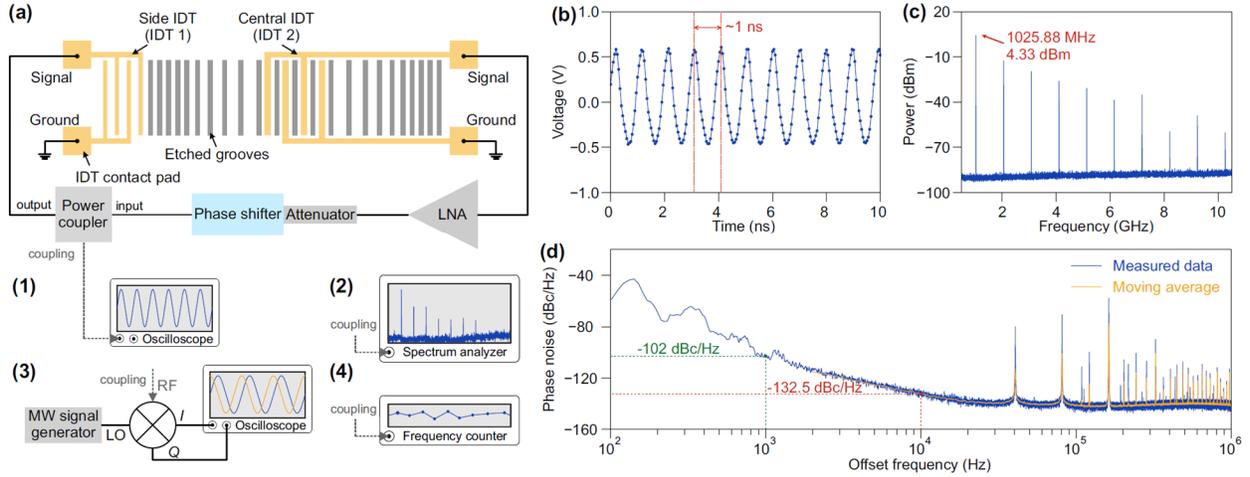

**FIG. 3. Experimental characterization of the surface acoustic wave (SAW) oscillator.** (a) Experimental setup. The SAW oscillator consists of the SAW resonator, a low noise amplifier (LNA), an attenuator, a microwave phase shifter and a microwave power coupler. The SAW oscillator outputs from the coupling port of the coupler. Insets (1)-(4) illustrate equipment used to characterize the oscillator output. Insets (1) and (2) are an oscilloscope and a spectrum analyzer, respectively. Inset (3), which consists of an in-phase/quadrature (I/Q) demodulator, a microwave signal generator and an oscilloscope, is used to measure the phase noise and the short-term frequency stability. Another ultra stable microwave source with frequency of 10s kHz lower than the oscillator feeds into the local oscillator (LO) port of the I/Q demodulator. Oscilloscope captures the I/Q data from the I/Q demodulator output ports for further data processing. Equipment in Inset (4) is a frequency counter which is used to characterize the long-term frequency stability. (b) Measured waveform, (c) frequency spectrum, and (d) phase noise of the SAW oscillator using equipment in Insets (1)-(3) of Fig. 3(a), respectively. The resolution bandwidth in (c) is set to 1 kHz.

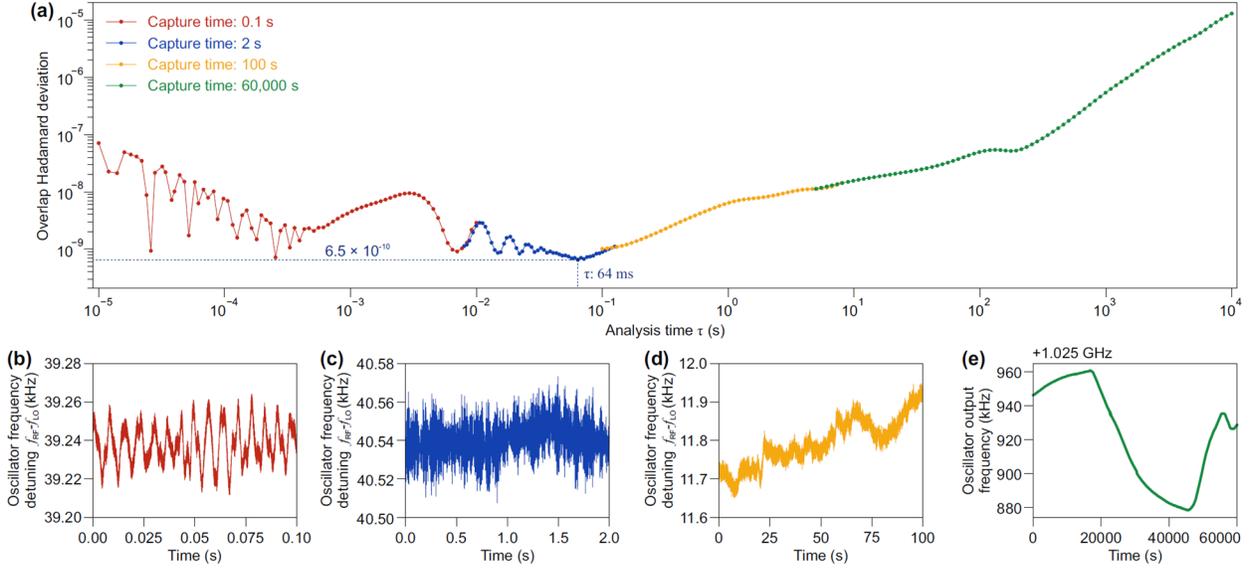

**FIG. 4. Long-term stability characterization of the oscillator**. (a) Measured overlapping Hadamard deviation of oscillator. Data of dot-solid line in red (the leftmost), blue and yellow color are calculated using the in-phase/quadrature (I/Q) data measured by the setup in Inset (3) of Fig. 3(a). Data in green color (the rightmost) is calculated using the data measured by the frequency counter (Inset (4) of Fig. 3(a)). (b), (c), (d) Extracted frequency from the measured I/Q signals in different time scales: (b) 0.1 sec, (c) 2 sec, and (d) 100 sec. (e) Measured frequency of the oscillator using a frequency counter over a long-term period of 60,000 seconds.

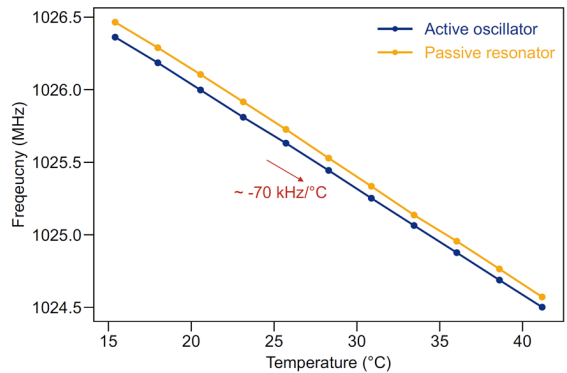

**FIG. 5. Frequency dependency on temperature of the SAW resonator and oscillator.** Blue (orange) data points represent the measured frequency of oscillator (the passive SAW resonator). Blue data points are the average frequency of 100-second measurements with a 100 µs sample interval. Orange data points are Lorentzian-fitted resonant frequencies of 10-times-averaged $S_{21}$ spectra measurements of the passive SAW resonator.

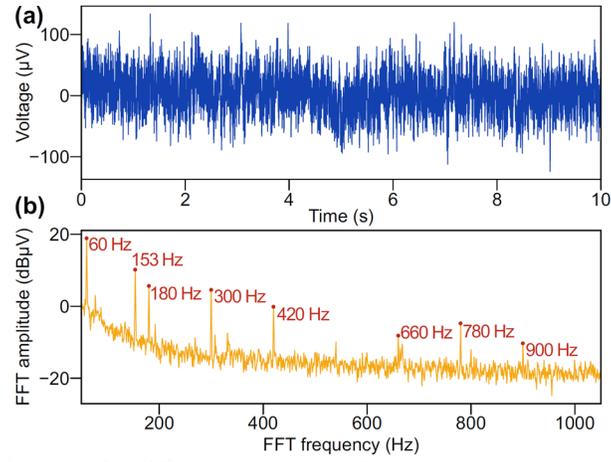

**FIG. 6.** (a) Measured waveform of the DC power supply. (b) Fourier transform of the measured waveform. The oscilloscope is set as AC-coupled.